\documentclass[twocolumn,showpacs,preprintnumbers,amsmath,amssymb]{revtex4}
\usepackage{graphicx}% Include figure files
\usepackage{dcolumn}% Align table columns on decimal point
\usepackage{bm}% bold math

\begin{document}

%\preprint{PRA 1 8 02 send}

\title{{\bf Quantum and Classical Noise in Practical Quantum Cryptography
Systems based on
polarization-entangled photons}}

\author{S. Castelletto}
 \altaffiliation[Permanent address: ]{Istituto Elettrotecnico Nazionale G.
Ferraris, Strada delle Cacce 91, 10135 Torino (Italy)}
  \email{castelle@ien.it}
\affiliation{%
Optical Technology Division\\
National Institute of Standards and Technology, Gaithersburg,
Maryland 20899-8441
}%

\author{I. P. Degiovanni}%
 \email{degio@ien.it}
\author{M. L. Rastello}

\affiliation{%
Istituto Elettrotecnico Nazionale G. Ferraris \\
Strada delle Cacce 91-10135 Torino (Italy)
}%

\date{\today}

\begin{abstract}
Quantum-cryptography key distribution (QCKD) experiments have been
recently reported using polarization-entangled photons. However,
in any practical realization, quantum systems suffer from either
unwanted or induced interactions with the environment and the
quantum measurement system, showing up as quantum and, ultimately,
statistical noise. In this paper, we investigate how ideal
polarization entanglement in spontaneous parametric downconversion
(SPDC) suffers quantum noise in its practical implementation as a
secure quantum system, yielding errors in the transmitted bit
sequence. Because all SPDC-based QCKD schemes rely on the
measurement of coincidence to assert the bit transmission between
the two parties, we bundle up the overall quantum and  statistical
noise in an exhaustive model to calculate the accidental
coincidences. This model predicts the quantum-bit error rate and
the sifted key and allows comparisons between different security
criteria of the hitherto proposed QCKD protocols, resulting in an
objective assessment of performances and advantages of different
systems.
\end{abstract}

\pacs{03.67.Dd, 03.65.Yz, 03.65.Ud, 42.65.-k}

\maketitle

\section{\label{sec:level1} Introduction}

Quantum Cryptography Key Distribution (QCKD) is at the moment the
most advanced and challenging application of quantum information.
QCKD offers the possibility that two remote parties, sender and
receiver (conventionally called Alice and Bob), can exchange a secret
random key, called sifted key (string of qubits), to implement a
secure encryption/decryption algorithm based on a shared secret
key, without the need that the two parties meet
\cite{bennet&brassard,ekert,bb92}.

In practical QCKD, Alice and Bob use a quantum channel, along
which sequences of signals are either sent or measured at random between
different bases of orthogonal quantum states. Alice can play the
role of either setting randomly the polarization basis of photons
and sending them to Bob (faint laser pulses as photon source), or
measuring photons randomly in any one of the selected bases
(entangled photon source). Bob, randomly and independently from
Alice, measures in one of the bases. The sifted key consists of
the subset of measurements performed when Alice's and Bob's bases
are in an agreed configuration according to the protocol used,
obtaining at this point a deterministic outcome whose security
relies on the laws of quantum physics, for they
previously agreed upon the correspondence between counting a
photon in a specific state and the bit values 0 or 1. In contrast,
the security of conventional cryptography relies upon
the unproven difficulty in
factorizing large numbers into prime numbers by a conventional
algorithm. We note that there is no guarantee that such an
algorithm does not exist.

The underlying feature of QCKD, namely the reliance of the
security of the distributed secret key on the laws of quantum
physics \cite{bennet&brassard,ekert,bb92}, gives it an advantage
over the public key cryptography. In other words, the uncertainty
principle prohibits one from gaining information from a quantum
channel without disturbing it. Basically, QCKD is founded on the
principle that when a third party (Eve) performs a measurement on
a qubit exchanged, she induces a perturbation, yielding errors in
the bit sequence transmitted, revealing her presence. Any attempt
by Eve to obtain information about the key leads to a nonzero
error rate in the generated sifted key. Nevertheless this last
claim must be somewhat softened because of practical realization
of quantum channels \cite{brassard}. Unfortunately, in practical
systems, errors also happen because of experimental leakage, like
losses in optics, detection, electronics and noise. Also, even
when no eavesdropper is disturbing the bit exchange, there will be
errors in the transmission and Alice's and Bob's strings will not
coincide perfectly. Thus in practice there is no way to
distinguish an eavesdropper attack from experimental
imperfections, making it necessary to establish an upper bound on
tolerable experimental imperfections in the realization of the
quantum channels  to implement an error correction procedure.

Following the first proposal by Bennett and Brassard
\cite{bennet&brassard} and later the Ekert protocol invoking
entangled states \cite{ekert}, various systems of QCKD have been
implemented and tested by groups around the world. Recently some
research groups \cite{ekertrarity,sasha,qk2,qk3,qk1} performed the
first QCKD experiments based on polarization-entangled photon
pairs, and Brassard \textit{et al}. \cite{brassard} proved
theoretically that QCKD schemes based on spontaneous parametric
down-conversion (SPDC) offer enhanced performance, mostly in terms
of security, compared to QCKD based on weak coherent pulses.

Entangled photons, generated by SPDC in nonlinear crystals, have
proved largely successful for quantum optical communication
\cite{mandel,rarity0,rarity,rarity2} and quantum radiometry
\cite{prl,klyshko,rarity4,peninserg,kwiat,migdall,metrolo,teichSaleh,josaB,madrideta}%
. Furthermore, basic experimental tests of the foundation of
quantum mechanics had been performed by exploiting the
entanglement of this source \cite{rarity3,gisin1,zeil}. However,
quantum noise in a SPDC quantum state may significantly limit
performance of the proposed quantum optical communication and
information technologies.

In this paper we provide a general model for an \textit{a priori}
evaluation of some crucial parameters of a general QCKD scheme
based on polarization entangled photons. We basically adopt the
formalism of quantum operations \cite{NC00} to describe the
dynamics of an open quantum system subject either to the
interactions with the environment or to a quantum device
performing a measurement on it. These unwanted or induced
interactions show up as noise in quantum-information processing
systems, degrading their ideal performance. Exploiting the
quantum-operation formalism we present a model to
quantify precisely both quantum and ultimately statistical noise
in quantum-information experiments performed using an entangled
photon source.

In Section II we consider quantum noise in a lossless measurement
system where noise is due to the coupling between the polarization
mode of the source with the polarizing beam splitter ports. In
Section III we discuss the case of a lossy system, where noise is
induced by detection deficiencies, such as losses of correlated
photons from the presence of optical elements, non-ideal
detectors, and electronic devices in either channel, and detector
dark-counts.

The model concludes with the calculation of an overall probability
of total coincidence counts, including an imperfect
time-correlation measurement, ultimately yielding an estimate of
accidental coincidences (Section IV).

This result is used in the calculation of the quantum bit error
rate (QBER) for QCKD protocols, i.e., BB84 and the two variants of
the Ekert's protocol, based on CHSH and Wigner's inequalities
respectively (Section V). Previsions are also presented about the
sifted and the corrected key, the quantum-bit error rate (QBER)
before and after a standard error correction procedure. Finally we
evaluate the performance of security criteria for Ekert's
protocols based on both the CHSH inequality and Wigner's
inequality (Section VI).

\section{\label{sec:level1} Quantum noise in polarization selection of
photon pairs}

In this Section we consider a real either non-maximally
entangled or partially mixed state resulting from both imperfect
entangled state generation by SPDC and imperfect polarization
state selection by real polarizing beam splitters (PBSs).

\begin{figure}[tbp]
%[htbp]
\par
\begin{center}
\includegraphics[angle=0, width=8 cm, height=5 cm]{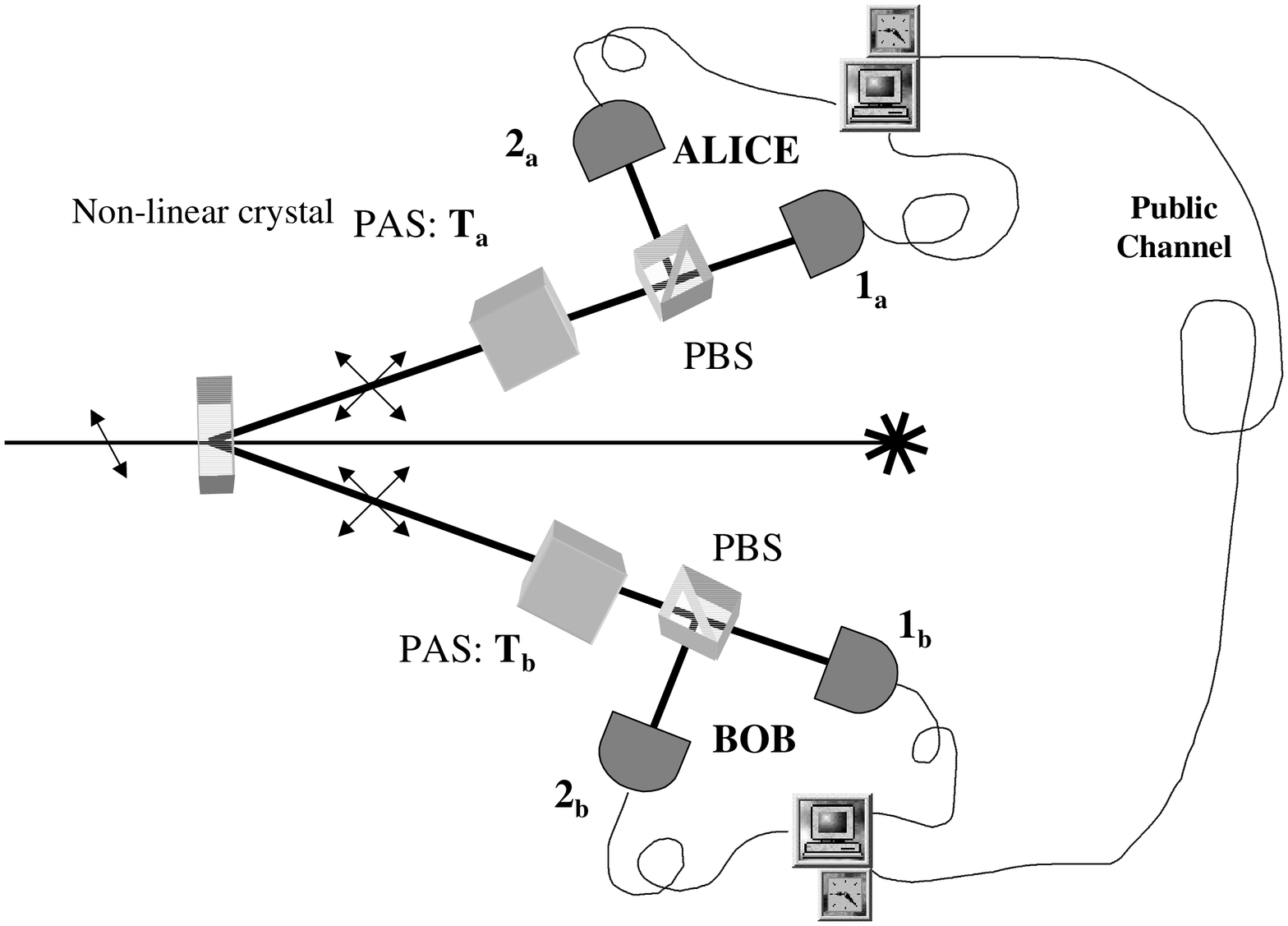}
\end{center}
\caption{ QCKD set-up: polarization-entangled photons generated by
SPDC\ are directed to the two parties (Alice and Bob). The bit
sequence of the key is obtained by means of polarization-sensitive
synchronized measurements performed by Alice and Bob according to
a specific QCKD protocol. } \label{Figure 1}
\end{figure}

In Fig. 1 we depict the typical scheme for quantum cryptography
key distribution as implemented using entangled photons
generated by SPDC. Because of a nonlinear interaction in a $%
\chi ^{(2)}$ crystal, some pump photons (angular frequency $\omega
_{p}$) spontaneously split into a lower-frequency pair of
photons, historically called signal and idler, perfectly
correlated in all aspects of their state (direction, energy,
polarization, under the constraints of conservation of energy and
wavevector momentum, otherwise known as phase-matching)$.$ These
entangled states show perfect correlation for polarization
measured along orthogonal but arbitrary axes.

The QCKD performed by pure entangled states relies on the
realization of two quantum correlated optical channels. These channels yield
single-photon polarization states, such that whenever Alice performs
a polarization measurement on a photon of the pair, automatically
the other photon is projected in a defined polarization state,
i.e., Alice plays the role of triggering Bob's measurement.
In actuality, the light field emerging from the output of the nonlinear
crystal is a polarization-entangled multimode state. However, it
can be described as a polarization entangled two-photon state in
only two effective modes (one for channel $a$ and the other for
channel $b$), as signal and idler pairs can be easily emitted
non-collinearly with the pump by proper phase-matching rules
\cite{opteng}. This scheme eventually is exploited in quantum
information applications by using one channel as the trigger or
reference ($a$) and the other channel as the probe ($b$). According to
Fig. 1, we denote Alice's detector apparatus to be the trigger
and Bob's to be the probe. Alice and Bob detection apparatus consist of
 polarization-analyzer systems (PAS) for proper single-beam
polarization rotation, polarizing beam splitters (PBS), photon detectors
($1_{a},\,2_{a},\,1_{b},\,2_{b}$), data storage systems
(computers) and synchronization systems.

Let us consider in the following type II SPDC entangled states
\cite{kwiat2}, where the output two-photon states are  a quantum
superposition of orthogonally polarized photons, i.e. the singlet
state \cite{foot1}:
\begin{equation*}
\left| \psi^{-} \right\rangle =\frac{1}{\sqrt{2}}\left( \left|
H_{a}\right\rangle \left| V_{b}\right\rangle -\left|
V_{a}\right\rangle \,\left| H_{b}\right\rangle \right) .
\end{equation*}

Practically, pure entanglement may not be achieved because of
imperfect source generation and because of incomplete entangled
photon collection. According to Refs. \cite{scienceK,berglund} an
uncompensated for coherence loss, induced in the state
by the coupling between polarization and frequency modes because
of a birefringent environment, may produce a partially mixed or
non-maximally entangled state.  Also the collection of the same number of
entangled
photons on both channels is unlikely, mostly due to the imperfect
positioning of the detection systems along the true directions of
entangled photons on the SPDC cones. Two complex variables, $\zeta
$\thinspace\ ($\left| \zeta \right| \leqslant 1$) and $\epsilon $,
characterize the imperfect compensation of dephasing and
decoherence in the crystal and the misalignment in collecting
entangled photons in the optical paths, respectevely. The net result is a
non-maximally entangled, or partially mixed state, written as
\begin{eqnarray*}
\widehat{\rho }^{\psi } &=&\frac{1}{1+\left| \epsilon \right| ^{2}} \\
&&\left[
\begin{array}{c}
\left| H_{a}\right\rangle \left| V_{b}\right\rangle \left\langle
V_{b}\right| \left\langle H_{a}\right| +\left| \epsilon \right|
^{2}\,\left| V_{a}\right\rangle \left| H_{b}\right\rangle
\left\langle H_{b}\right|
\left\langle V_{a}\right| + \\
-\epsilon \zeta \left| V_{a}\right\rangle \left|
H_{b}\right\rangle \left\langle V_{b}\right| \left\langle
H_{a}\right| -(\epsilon \zeta )^{\ast }\left| H_{a}\right\rangle
\left| V_{b}\right\rangle \left\langle H_{b}\right| \left\langle
V_{a}\right|
\end{array}
\right] .
\end{eqnarray*}
We analyze now the quantum noise introduced via an imperfect
polarization state selection by the PBSs depicted in Fig. 1. Here,
the entangled photons are detected accordingly to their
polarizations by using imperfect polarizing beam splitters (PBSs)
on both arms and perfect single-photon detectors. The PBSs project
photons onto a polarization basis $\left\{ \left| H_{a}\right\rangle
\left| H_{b}\right\rangle ,\left| V_{a}\right\rangle \left|
V_{b}\right\rangle ,\left| H_{a}\right\rangle \left|
V_{b}\right\rangle ,\left| V_{a}\right\rangle \left|
H_{b}\right\rangle \right\} ,$ while the polarization analyzer
systems (PASs) induce the
transformations represented by the unitary operators $\widehat{T}_{z}$  in
each
arm ($z=a,b$), according to
\begin{eqnarray*}
\widehat{T}_{z}\left| H_{z}\right\rangle &=&c_{1}\left|
H_{z}\right\rangle
+c_{2}\left| V_{z}\right\rangle , \\
\widehat{T}_{z}\left| V_{z}\right\rangle &=&c_{2}\left|
H_{z}\right\rangle -c_{1}\left| V_{z}\right\rangle .
\end{eqnarray*}

In the case of ideal PBSs, channel $z$ transmits state $\left|
H_{a}\right\rangle $ ($\left| H_{b}\right\rangle $) and reflects
state $\left| V_{a}\right\rangle $ ($\left| V_{b}\right\rangle $),
i.e., there is a perfect coupling between the output ports of the
PBSs and the projections of the photon polarization state. In the
approach so far adopted in the literature a perfect coupling is
always assumed because the
measurement process is considered as a projection on polarization states $%
\left| H_{a}\right\rangle ,$ $\left| V_{a}\right\rangle $,\
$\left|
H_{b}\right\rangle ,$ $\left| V_{b}\right\rangle ,$ thus assuming detectors $%
1_{a},\,1_{b},2_{a}\,,2_{b},$ are sensitive to polarization$.$ Here we
consider a further noise effect induced by the presence of real
PBSs, where a small part of the photons projected onto $\left|
V_{a}\right\rangle $ $%
(\left| V_{b}\right\rangle )$ are erroneously transmitted, and
some photons projected in the state $\left| H_{a}\right\rangle $
$(\left| H_{b}\right\rangle )$ are erroneously reflected
(generally less than 2 \% and 5 \%, respectively), as shown
in Fig. 2. For this purpose, we extend the Hilbert space to
describe these photon states as
\begin{eqnarray}
\left| H_{z}\right\rangle \left| O1_{z}\right\rangle
&=&t_{z}\left| H_{z}\right\rangle \left| I1_{z}\right\rangle
+r_{z}\left|
H_{z}\right\rangle \left| I2_{z}\right\rangle  \notag \\
\left| H_{z}\right\rangle \left| O2_{z}\right\rangle
&=&t_{z}\left| H_{z}\right\rangle \left| I2_{z}\right\rangle
+r_{z}\left|
H_{z}\right\rangle \left| I1_{z}\right\rangle  \notag \\
\left| V_{z}\right\rangle \left| O1_{z}\right\rangle
&=&r_{z}^{\perp }\left| V_{z}\right\rangle \left|
I2_{z}\right\rangle +t_{z}^{\perp }\left|
V_{z}\right\rangle \left| I1_{z}\right\rangle  \notag \\
\left| V_{z}\right\rangle \left| O2_{z}\right\rangle
&=&r_{z}^{\perp }\left| V_{z}\right\rangle \left|
I1_{z}\right\rangle +t_{z}^{\perp }\left| V_{z}\right\rangle
\left| I2_{z}\right\rangle ,  \label{eq5}
\end{eqnarray}
where $\left| Ox_{z}\right\rangle $ represents the photon crossing
the output port of the PBS towards detector $x_{z}$ and $\left|
Ix_{z}\right\rangle $ represents the photon crossing the input
port of the PBS, with $x=1,2$. $\left| t_{z}^{\perp }\right| ^{2}$
is the transmittance
of photons in the $\left| V_{z}\right\rangle $ polarization state, and $%
\left| r_{z}^{\perp }\right| ^{2}=1-$\ $\left| t_{z}^{\perp
}\right| ^{2}$ is the corresponding reflectance, whose phase
relation is $r_{z}^{\perp }/t_{z}^{\perp }=i\left| r_{z}^{\perp
}\right| /\left| t_{z}^{\perp }\right| $. Analogously, $\left|
t_{z}\right| ^{2}$ and $\left| r_{z}\right| ^{2}$ are the
transmittance and reflectance of photons in the $\left|
H_{z}\right\rangle $ state. So far, we have considered only
lossless PBSs. The effect of photon losses due to all
optical devices used in the channels are treated in Sections III
and IV.

\begin{figure}[tbp]
%[htbp]
\par
\begin{center}
\includegraphics[angle=0, width=8 cm, height=5 cm]{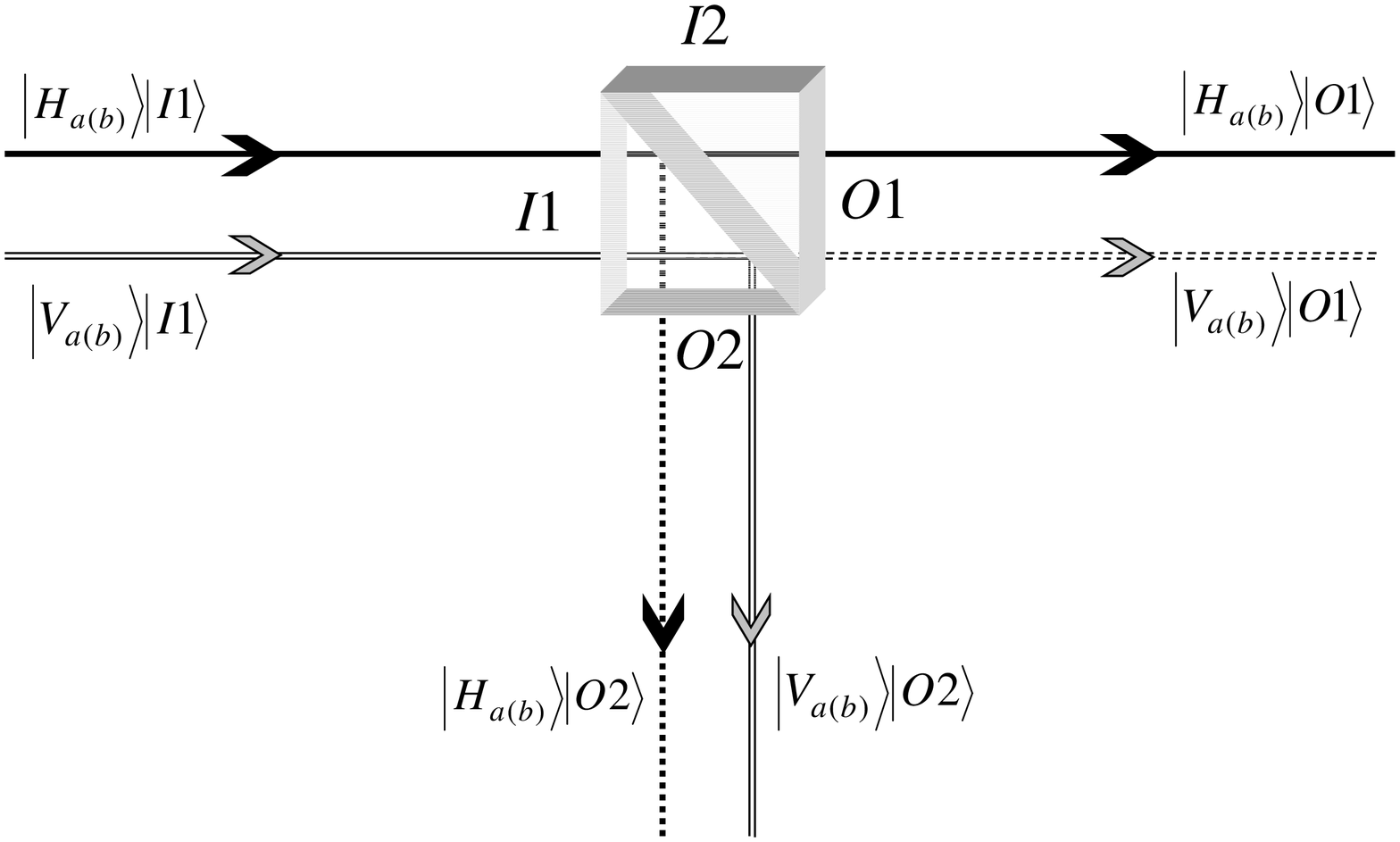}
\end{center}
\caption{ Real PBS: all photons projected in $\left|
V_{a(b)}\right\rangle $ polarized state should be reflected, but
some of these are wrongly transmitted. Moreover all photons
projected in $\left| H_{a(b)}\right\rangle $ polarized state
should be transmitted, but some of these are erroneously
reflected. } \label{Figure 2}
\end{figure}

We define the input density matrix for the PBS ports as
\begin{equation*}
\widehat{\rho }^{\text{PBS}}=\left| I1_{a}\right\rangle \left|
I1_{b}\right\rangle \left\langle I1_{b}\right| \left\langle
I1_{a}\right| ,
\end{equation*}
and the total input density matrix $\widehat{\rho }^{\text{in}}$
of the photon system as
\begin{equation*}
\widehat{\rho }^{\text{in}}= \widehat{%
\rho }^{\psi } \otimes \widehat{\rho }^{\text{PBS}} .
\end{equation*}

The formalism of quantum operation is the most suitable to
describe the evolution of a quantum system coupled with another
quantum system or with the environment, as well as the evolution of
a quantum system subject to measurement \cite{NC00}. In this
context, we consider the set of non-trace-preserving quantum
operations $\left\{ \mathcal{E}_{x_{a}y_{b}}\right\} $ defined as
\[
\mathcal{E}_{x_{a}y_{b}}\left( \widehat{\rho }^{\psi }\right) =\text{Tr}_{%
\text{PBS}}\left( \widehat{P}_{x_{a}y_{b}}\widehat{U}_{\psi \text{-PBS}}%
\widehat{S}_{\psi }\widehat{\rho }^{\text{in}}\widehat{S}_{\psi }^{\dagger }%
\widehat{U}_{\psi \text{-PBS}}^{\dagger
}\widehat{P}_{x_{a}y_{b}}^{\dagger }\right) ,
\]
describing the process of detection of the photon pair by the detectors
$x_{a}$ and $y_{b}$ ($x,y=1,2$). In this
expression, the unitary operator $\widehat{S}_{\psi }$ describes
the action of the PASs and the unitary transformation
$\widehat{U}_{\psi \text{-PBS}}$ describes the coupling between the photon
pair polarization state and the PBS ports. The explicit form of
$\widehat{U}_{\psi \text{-PBS}}$ is deduced by Eq.s (\ref
{eq5}), and calculations are reported in Appendix A. Because the operators $%
\widehat{T}_{a}$ and $\widehat{T}_{b}$ independently act on the
corresponding subspaces $a$ and $b$ of the Hilbert space of
polarization and induce linear transformation, they are
ineffective on the Hilbert space of the PBS ports. Thus, $\widehat{\rho
}^{\text{in}}$ is subject to a global
transformation written as an unitary operator $\widehat{S}_{\psi }=\mathbf{1}%
^{\text{PBS}}\otimes \widehat{S}_{a,b},$ where \ $\widehat{S}_{a,b}=\widehat{%
T}_{a}\otimes \widehat{T}_{b}.$

$\widehat{P}_{x_{a}y_{b}}=\left| Ox_{a}\right\rangle \left|
Oy_{b}\right\rangle \left\langle Oy_{b}\right| \left\langle
Ox_{a}\right| $
is the projector representing the detection process by the two detectors $%
x_{a}$ and $y_{b}.$ The probability of detection of the photon
pair by the detectors $x_{a}$ and $y_{b}$ is

\begin{eqnarray}
p(x_{a},y_{b}) &=&\text{Tr}_{\psi }\left[
\mathcal{E}_{x_{a}y_{b}}\left(
\widehat{\rho }^{\psi }\right) \right] =  \notag \\
&=&\text{Tr}\left( \widehat{P}_{x_{a}y_{b}}\widehat{U}_{\psi \text{-PBS}}%
\widehat{S}_{\psi }\widehat{\rho }^{\text{in}}\widehat{S}_{\psi }^{\dagger }%
\widehat{U}_{\psi \text{-PBS}}^{\dagger
}\widehat{P}_{x_{a}y_{b}}^{\dagger }\right). \label{eqste}
\end{eqnarray}
$\sum_{x_{a}y_{b}}\mathcal{E}_{x_{a}y_{b}} \left( \widehat{\rho
}^{\psi }\right)$ is trace-preserving because the probabilities of
the distinct outcomes sum
to one, i.e., Tr$_{\psi }\left[ \sum_{x_{a}y_{b}}\mathcal{E}%
_{x_{a}y_{b}}\left( \widehat{\rho }^{\psi }\right) \right]
=\sum_{x_{a}y_{b}}p(x_{a},y_{b})=1$ for all possible input $\widehat{\rho }%
^{\psi }.$

\section{ Quantum and classical noise in photon counts}

In the following we consider the noise contribution to the photon
counts because of an imperfect collection of photons and a
noisy and lossy detection system. For the experimental setup in
Fig. 1, we calculate the total probability
$p_{\text{tot},x_{z}}(n)$ of $n$ counts by any detector $x_{z}$ by
separately calculating the probabilities of counts associated with
correlated photons ($p_{sp,x_{z}}(n)$), with uncorrelated photons
($p_{u,x_{z}}(n)$), and with detector dark counts
($p_{d,x_{z}}(n)$).

To describe the counting process we adopt the formalism
of quantum operations, where we consider a general density matrix
representing photons on a channel $\nu$ in term of number of
photons, i.e. $\left\{ \left| n^{\nu}\right\rangle \right\} ,$ as
\begin{equation}
\widehat{\rho }^{\nu}=\sum_{n,m=0}^{\infty }\rho
_{\,nm}^{\nu}\left| n^{\nu}\right\rangle \left\langle
m^{\nu}\right| .  \label{rhozeta}
\end{equation}
The evolution of the system $\widehat{\rho }^{\nu}$ is evaluated according
to the formalism of quantum operations.

In this way, we define the set of non-trace-preserving quantum
operations as $\left\{ \mathcal{E}_{m}^{\mu }\right\} $ as
\begin{equation}
\mathcal{E}_{m}^{\mu }\left( \widehat{\rho }^{\nu }\right) =\text{Tr}_{\text{%
E}_{\mu }}\left( \widehat{P}_{m}^{\mu }\widehat{U}_{\text{Q}_{\nu }\text{-E}%
_{\mu }}\widehat{\rho }^{\nu }\otimes \left| e_{0}^{\mu
}\right\rangle \left\langle e_{0}^{\mu }\right|
\widehat{U}_{\text{Q}_{\nu }\text{-E}_{\mu }}^{\dagger
}\widehat{P}_{m}^{\mu \;\dagger }\right) ,  \label{exz}
\end{equation}
which describes the detection of $m$ photons by the system $\mu $. In
this expression the unitary operator $\widehat{U}_{\text{Q}_{\nu }\text{-E}%
_{\mu }}$  represents the interaction between the quantum system
Q$_{\nu }$ of photons in the channel $\nu $ in the
initial state $\widehat{\rho }^{\nu }$ and the lossy and noisy environment E$%
_{\mu }$ in the initial state\ $\left| e_{0}^{\mu }\right\rangle
$. The action of $\widehat{U}_{\text{Q}_{\nu }\text{-E}_{\mu }}$
on the state ``number of photons'' is
\begin{equation}
\widehat{U}_{\text{Q}_{\nu }\text{-E}_{\mu }}\left| n^{\nu
}\right\rangle \left| e_{0}^{\mu }\right\rangle
=\sum_{m=0}^{n}d^{\mu }(m,n)\left| m^{\nu }\right\rangle \left|
e_{m,n}^{\mu }\right\rangle  \label{uxz}
\end{equation}
where $\left| d^{\mu }(m,n)\right| ^{2}$ is the probability of
measuring $m$ photons out of $n$ present in the channel $\nu $
because of losses. $\widehat{P}_{m}^{\mu }=\sum_{n=m}^{\infty
}\left| e_{m,n}^{\mu }\right\rangle \left\langle e_{m,n}^{\mu
}\right| $ are the measurement operators.

Thus, the probability of measuring $m$ counts by the detection
system $\mu $ is
\begin{equation}
p_{\mu }(m)=\text{Tr}_{\text{Q}_{\nu }}\left[ \mathcal{E}_{m}^{\mu
}\left( \widehat{\rho }^{\nu }\right) \right] =\sum_{n=m}^{\infty
}\rho _{\,nn}^{\nu }\left| d^{\mu }(m,n)\right| ^{2}  \label{pisp}
\end{equation}
where $\rho _{\,nn}^{\nu }$ is the probability of having $n$
photons on the channel $\nu$.

\subsection{Single counts associated with correlated photons}

In the following we concentrate primarily on the measurement of
correlated photons by a lossy detector, $x_{z}$. We define the density
matrix of the number of
photon pairs, $\widehat{\rho }^{p}$, as
\begin{equation}
\widehat{\rho }^{p}=\sum_{n,m=0}^{\infty }\rho _{\,nm}^{p}\left|
n^{p}\right\rangle \left\langle m^{p}\right| .  \label{rhopi}
\end{equation}
Analogous to Eq. (\ref{rhopi}), we write a density matrix
of single photons of the pairs (sp) along channel $z$:
\begin{equation}
\widehat{\rho }^{sp,z}=\sum_{n,m=0}^{\infty }\rho
_{\,nm}^{p}\left| n^{sp,z}\right\rangle \left\langle
m^{sp,z}\right| .
\end{equation}

The counting of correlated photons on channel $z$ by the
detector $x_{z}$ is described by mapping $\widehat{\rho
}^{sp,z}$ on the set of non-trace-preserving quantum operation $\left\{
\mathcal{E}%
_{m}^{sp,x_{z}}\right\} $.
The explicit form of $\mathcal{E}_{m}^{sp,x_{z}}\left( \widehat{\rho }%
^{sp,z}\right) $ is deduced by analogy to Eq. (\ref{exz}), by
replacing the interaction unitary operator with
$\widehat{U}_{\text{Q}_{sp,z}\text{-E}%
_{sp,x_{z}}}$ and the measurement operator with $\widehat{P}%
_{m}^{sp,x_{z}}$, given that Q$_{sp,z}$ is the quantum system of
single photons of the pair on the channel $z$ and E$%
_{sp,x_{z}}$ is the lossy and noisy environment in the initial
state\ $\left| e_{0}^{sp,x_{z}}\right\rangle $. The
$\widehat{U}_{\text{Q}_{sp,z}\text{-E}_{sp,x_{z}}}$ action on the
state $\left| n^{sp,z}\right\rangle $ is completely described by
means of coefficients $d^{sp,x_{z}}(m,n)$ in complete analogy with
Eq. (\ref{uxz}), while we have
$\widehat{P}_{m}^{sp,x_{z}}=\sum_{n=m}^{\infty }\left|
e_{m,n}^{sp,x_{z}}\right\rangle \left\langle
e_{m,n}^{sp,x_{z}}\right| $.

The probability of $m$ counts  by the detector $x_{z}$ corresponding to
correlated photons
 becomes
\begin{equation}
p_{sp,x_{z}}(m)=\sum_{n=m}^{\infty }\rho _{\,nn}^{p}\left|
d^{sp,x_{z}}(m,n)\right| ^{2}. \label{pisp}
\end{equation}
The probability of $n$ pairs is given by $\rho _{\,nn}^{p}=\left( \,\lambda
_{p}t\right) ^{n}\,\exp (-\lambda _{p}t)/n!,$ $t$ being the time
of measurement and $\lambda _{p}$ the mean rate of photon pairs in the
Alice and Bob channels \cite{teichSaleh,capri,Perina,josab2}. The
terms
\begin{eqnarray*}
\left| d^{sp,x_{a}}(m,n)\right| ^{2} &=&\left(
\begin{array}{c}
n \\
m
\end{array}
\right) \left[ \xi _{x_{a}}\sum_{y_{b}}p(x_{a},y_{b})\right] ^{m} \\
&&\left[ 1-\xi _{x_{a}}\sum_{y_{b}}p(x_{a},y_{b})\right] ^{n-m}, \\
\left| d^{sp,y_{b}}(m,n)\right| ^{2} &=&\left(
\begin{array}{c}
n \\
m
\end{array}
\right) \left[ \xi _{y_{b}}\sum_{x_{a}}p(x_{a},y_{b})\right] ^{m} \\
&&\left[ 1-\xi _{y_{b}}\sum_{x_{a}}p(x_{a},y_{b})\right] ^{n-m}.
\end{eqnarray*}
are the probabilities that only $m$ out of $n$ photons in the
channel $a$ ($b$) are counted by the detector $x_{a}$ ($y_{b}$).
Losses due to electronics ($\pi _{z}$),  detection efficiencies
($\eta _{x_{z}}$), as well as optical losses ($\tau _{x_{z}}$) are
summed up in the term $\xi _{x_{z}}=\pi _{z}\eta _{x_{z}}\tau
_{x_{z}}$ \cite{josaB,madrid}$;$ while we refer to the Appendix B
for the analysis of dead time in this context. The term $\tau
_{x_{z}}$ incorporates all losses in the Alice and Bob optical
path, such as from crystals,
filters, lenses, PBSs, PASs and fibers. The terms $%
\sum_{x_{a}(y_{b})}p(x_{a},y_{b})$ are the probability that each
photon of the pair may be counted randomly by any arbitrary
detector (Eq. \ref{eqste}). Probability $p_{sp,x_{z}}(m)$ is
derived according to Eq. (\ref{pisp}), giving
\begin{equation}
p_{sp,x_{z}}(n)=\left( \lambda _{sp,x_{z}}t\right) ^{n}\frac{\exp
\left( -\lambda _{sp,x_{z}}t\right) }{n!},  \label{pisp2}
\end{equation}
with mean count rates given by
\begin{eqnarray*}
\lambda _{sp,x_{a}} &=&\xi _{x_{a}}\sum_{y_{b}}p(x_{a},y_{b})\lambda _{p}, \\
\lambda _{sp,y_{b}} &=&\xi
_{y_{b}}\sum_{x_{a}}p(x_{a},y_{b})\lambda _{p}.
\end{eqnarray*}

\subsection{Single counts associated with uncorrelated photons and dark counts}

Here we consider counts from any detector $x_{z}$ from stray light,
uncorrelated photons, and dark counts eventually contributing to
noise in the distributed key. The density matrix associated with
stray light and uncorrelated photons is
\begin{equation*}
\widehat{\rho }^{u,x_{z}}=\sum_{n,m=0}^{\infty }\rho
_{\,nm}^{u,x_{z}}\left| n^{u,x_{z}}\right\rangle \left\langle
m^{u,x_{z}}\right| .
\end{equation*}
By pursuing the same formalism as before the detection of
uncorrelated photons by the detector $x_{z}$  is described by
means of the set of non-trace-preserving quantum operation
$\left\{ \mathcal{E}_{m}^{u,x_{z}}\right\} .$ The map
$\mathcal{E}_{m}^{u,x_{z}}\left( \widehat{\rho }^{u,x_{z}}\right)
$ follows in analogy with Eq. (\ref{exz}). The unitary operator $\widehat{U}%
_{\text{Q}_{u,x_{z}}\text{-E}_{u,x_{z}}}$ describes the
interaction between the quantum system, Q$_{u,x_{z}}$, of
uncorrelated photons on the channel $x_{z}$ and the lossy
environment, E$_{u,x_{z}}$, in the initial state\ $\left|
e_{0}^{u,x_{z}}\right\rangle $. The measurement operator is
$\widehat{P}_{m}^{u,x_{z}}=\sum_{n=m}^{%
\infty }\left| e_{m,n}^{u,x_{z}}\right\rangle \left\langle
e_{m,n}^{u,x_{z}}\right| .$
The action of $\widehat{U}_{\text{Q}_{u,x_{z}}%
\text{-E}_{u,x_{z}}}$ on the state $\left|
n^{u,x_{z}}\right\rangle $ follows from Eq.
(\ref{uxz}) with the decomposition coefficients,
$d^{u,x_{z}}(m,n)$.

Thus, the probability of measuring $m$ counts of uncorrelated
photons by the detector $x_{z}$ is
\begin{equation}
p_{u,x_{z}}(m)=\sum_{n=m}^{\infty }\rho _{\,nn}^{u,x_{z}}\left|
d^{u,x_{z}}(m,n)\right| ^{2}, \label{piu}
\end{equation}
where $\rho _{\,nn}^{u,x_{z}}$ is the probability of $n$
uncorrelated photons in the channel $x_{z}$. According to Refs.
\cite{teichSaleh,capri,Perina,josab2}, we assume that we have
$\rho _{\,nn}^{u,x_{z}}=\left( \lambda _{u,x_{z}}t\right) ^{n}\exp
\left( -\lambda _{u,x_{z}}t\right) /n!,$ where $\lambda
_{u,x_{z}}$ is the mean rate of uncorrelated photons. The term
\begin{equation*}
\left| d^{u,x_{z}}(m,n)\right| ^{2}=\left(
\begin{array}{c}
n \\
m
\end{array}
\right) \left( \xi _{x_{z}}\right) ^{m}\left( 1-\xi
_{x_{z}}\right) ^{n-m}
\end{equation*}
is the probability of $m$ out of $n$ uncorrelated photons counted
by the detector $x_{z}$.

The $p_{u,x_{z}}(n)$ derived accordingly from Eq. (\ref{piu}) is
\begin{equation}
p_{u,x_{z}}(n)=\left( \xi _{x_{z}}\lambda _{u,x_{z}}t\right)
^{n}\frac{\exp \left( -\xi _{x_{z}}\lambda _{u,x_{z}}t\right)
}{n!}.
\end{equation}

The main source of noise in detectors is due to dark counts,
whose distribution is regarded merely from a statistical point of
view as the probability of $n$ dark counts
\begin{equation*}
p_{d,x_{z}}(n)=\left( \,\lambda _{d,x_{z}}t\right)
^{n}\,\frac{\exp (-\lambda _{d,x_{z}}t)}{n!},
\end{equation*}
with the mean dark-count rate being $\,\,\lambda _{d,x_{z}}$.

\subsection{Total counts}

As  real counters cannot distinguish among counts due to
correlated photons, counts due to uncorrelated photons, and dark
counts, the total probability of
measuring $k$ counts by detector $x_{z}$ is calculated according to
\cite{teichSaleh,capri,josab2}%
,
\begin{equation*}
p_{\text{tot},x_{z}}(k)=\sum_{l,m,n=0}^{\infty }\delta
_{k,l+m+n}\,p_{sp,x_{z}}(l)\,p_{u,x_{z}}(m)\,p_{d,x_{z}}(n),
\end{equation*}
giving
\[p_{\text{tot},x_{z}}(n)=\left( \,\lambda _{\text{tot}
,x_{z}}t\right) ^{n}\frac{\exp (-\lambda
_{\text{tot},x_{z}}t)}{n!},
\]
where the
mean rate of total counts measured by the detector $x_{z}$ is $\lambda _{%
\text{tot},x_{z}}=\lambda _{sp,x_{z}}+\xi _{x_{z}}\lambda
_{u,x_{z}}+\lambda _{d,x_{z}}.$

\section{\label{sec:level1} Coincidence counts}

We build up a model for the probability $p_{c,x_{a}y_{b}}(n)$ of measuring
$n$ coincidences by a pair of detectors $x_{a}$ and $%
y_{b}$ in order to estimate crucial quantities of a typical QCKD
experiment, such as the sifted key and the QBER before and after the
error correction procedure, whenever different protocols are
applied. We distinguish between the probability distribution of
true coincidences ($p_{p,x_{a}y_{b}}(n),$ due to correlated
photons) and the probability distribution of accidental
coincidences ($p_{Acc,x_{a}y_{b}}(n)$, because of imperfections in
the detection electronics).

We consider the density matrix in terms of counted pair states
(Eq. (\ref {rhopi})), and we describe its evolution exploiting the
formalism of quantum operations as described in Section IV by defining
another set of non-trace-preserving quantum operations $\left\{
\mathcal{E}_{m}^{p,x_{a}y_{b}}\right\} .$

$\mathcal{E}_{m}^{p,x_{a}y_{b}}\left( \widehat{\rho }^{p}\right) $
describes the measurement of $m$ coincidences originated by the
detection of the two photons of a pair by the detectors $x_{a}$, $%
y_{b}$. Its explicit expression is found from Eq. (\ref{exz}%
), except for the interaction between the quantum system Q$_{p}$
of photon pairs in the initial state $\widehat{\rho }^{p}$ and the lossy
and noisy environment E%
$_{p,x_{a}y_{b}}$ in the initial state\ $\left|
e_{0}^{p,x_{a}y_{b}}\right\rangle $ represented by the unitary operator $%
\widehat{U}_{\text{Q}_{p}\text{-E}_{p,x_{a}y_{b}}}$ and the measurement
operator $\widehat{P}%
_{m}^{p,x_{a}y_{b}}=\sum_{n=m}^{\infty }\left|
e_{m,n}^{p,x_{a}y_{b}}\right\rangle \left\langle
e_{m,n}^{p,x_{a}y_{b}}\right| .$

The action of $\widehat{U}_{\text{Q}_{p}%
\text{-E}_{p,x_{a}y_{b}}}$ on the state ``number of photon pairs''
is
\begin{equation*}
\widehat{U}_{\text{Q}_{p}\text{-E}_{p,x_{a}y_{b}}}\left|
n^{p}\right\rangle \left| e_{0}^{p,x_{a}y_{b}}\right\rangle
=\sum_{m=0}^{n}d^{p,x_{a}y_{b}}(m,n)\left| m^{p}\right\rangle
\left| e_{m,n}^{p,x_{a}y_{b}}\right\rangle .
\end{equation*}

Thus, the probability of measuring $m$ true coincidences
corresponding to photon pairs by the pair of detectors $x_{a},$
$y_{b}$ is
\begin{equation}
p_{p,x_{a}y_{b}}(m)=\sum_{n=m}^{\infty }\rho _{\,nn}^{p}\left|
d^{p,x_{a}y_{b}}(m,n)\right| ^{2}. \label{pipi}
\end{equation}

Realizing that a true coincidence may occur only if both photons
of the pair are not lost, we emphasize that the terms $\left|
d^{p,x_{a}y_{b}}(m,n)\right| ^{2}$ are the probabilities that only
$m$ pairs are detected as coincidences by the pair of detectors
$x_{a}$, $y_{b}$ when $n $ photons are present in the Alice's and
Bob's channels. It is straightforward to deduce the explicit form
of $\left| d^{p,x_{a}y_{b}}(m,n)\right| ^{2}$ as
\begin{eqnarray*}
\left| d^{p,x_{a}y_{b}}(m,n)\right| ^{2} &=&\left(
\begin{array}{c}
n \\
m
\end{array}
\right) \left[ \xi _{x_{a}}\xi _{y_{b}}p(x_{a},y_{b})\right] ^{m} \\
&&\left[ 1-\xi _{x_{a}}\xi _{y_{b}}p(x_{a},y_{b})\right] ^{n-m}.
\end{eqnarray*}
Probability $%
p_{p,x_{a}y_{b}}(n)$ is derived according to Eq. (\ref{pipi}),
obtaining
\begin{equation*}
p_{p,x_{a}y_{b}}(n)=\left( \lambda _{p,x_{a}y_{b}}t\right)
^{n}\frac{\exp \left( -\lambda _{p,x_{a}y_{b}}t\right) }{n!}\,
\end{equation*}
where $\lambda _{p,x_{a}y_{b}}=\xi _{x_{a}}\xi
_{y_{b}}p(x_{a},y_{b})\lambda _{p}$ is the mean rate of true coincidences
seen by the pair of detectors $x_{a}$ and $%
y_{b}$.

\subsection{Accidental coincidences \protect\bigskip}

The presence of the temporal coincidence window$\,w,$ during which
coincidences are measured, modifies the mean total coincidence
counts, thus forcing one to distinguish between true and
accidental coincidence statistics. We assume that true
coincidences occur in the middle of the coincidence temporal
window. Then we deduce the probability distribution of accidental
coincidences and finally the probability distribution of total
coincidences$, $ accounting for true and accidental coincidences,
assuming $w<D_{z}$, where $D_{z}\ $is the dead time in the $z$
channel according to Appendix B.

We regard $p_{\text{N},x_{z}}(n)$ as the probability distribution
of photons counted by the detector $x_{z}$ that may contribute to
accidental coincidences in the time interval $\Delta t$. By
observing that the probability distributions
$p_{\text{tot},x_{z}}(n)$ and $p_{p,x_{a}y_{b}}(n)$ are
Poisson, it is simple to demonstrate that we have
\begin{equation*}
p_{\text{N},x_{z}}(n)=\left( \lambda _{\text{N},x_{z}}\Delta t\right) ^{n}%
\frac{\exp \left( -\lambda _{\text{N},x_{z}}\Delta t\right) }{n!},
\end{equation*}
where $\lambda _{\text{N},x_{a}}=\lambda
_{\text{tot},x_{a}}-\lambda
_{p,x_{a}y_{b}}$, and $\lambda _{\text{N},y_{b}}=\lambda _{\text{tot}%
,y_{b}}-\lambda _{p,x_{a}y_{b}}$ are the mean count rates possibly
contributing to accidental coincidences from the detectors $x_{a}$
and $y_{b}$, respectively.

Let us denote by $q_{y_{b}}$ the probability that at least one
photon in $y_{b}$ is counted in the coincidence window $\Delta
t=w$
\begin{equation}
q_{y_{b}}=\sum_{n=1}^{\infty }p_{\text{N},x_{z}}^{\Delta
t=w}(n)=1-\exp \left( -\lambda _{\text{N},x_{z}}w\right) ,
\label{eqq}
\end{equation}
because detectors $x_{a}$ are here considered as triggers.

The term in Eq. (\ref{eqq}) is intended to account for the
contribution of single detectors $y_{b}$. Because detectors
$y_{b}$ are statistically independent, the probability that both
detectors count a photon producing an accidental coincidence is
$q_{1_{b}}\,q_{2_{b}}$. The final probability of accidental counts
from detector $y_{b}$ is obtained by subtracting half the
probability that both detectors in Bob's channel count an
accidental photon, in formula
\begin{eqnarray}
q_{1_{b}}^{\ast } &=&q_{1_{b}}(1-\frac{1}{2}q_{2_{b}}),  \label{q1*} \\
q_{2_{b}}^{\ast } &=&q_{2_{b}}(1-\frac{1}{2}q_{1_{b}}).
\label{q2*}
\end{eqnarray}

According to \cite{teichSaleh,capri,Perina,josab2}, we calculate
the
probability distribution of accidental coincidences in the time measurement $%
t$ by\ applying the discrete convolution between the Poisson
distribution of ``triggering'' counts and the binomial
distribution with parameter $q_{y_{b}}^{\ast },$
\begin{equation*}
p_{Acc,x_{a}y_{b}}(m)=\sum_{n=m}^{\infty }p_{\text{N},x_{a}}(n)B_{y_{b}}%
\left( m,n\right)
\end{equation*}
with $B_{y_{b}}\left( m,n\right) =$\bigskip $\left(
\begin{array}{c}
n \\
m
\end{array}
\right) \left( q_{y_{b}}^{\ast }\right) ^{m}\left(
1-q_{y_{b}}^{\ast }\right) ^{n-m},$ giving
\begin{equation*}
p_{Acc,x_{a}y_{b}}(n)=\left( \lambda _{Acc,x_{a}y_{b}}t\right) ^{n}\frac{%
\exp \left( -\lambda _{Acc,x_{a}y_{b}}t\right) }{n!}
\end{equation*}
with $\lambda _{Acc,x_{a}y_{b}}=q_{y_{b}}^{\ast }\lambda
_{\text{N},x_{a}}.$

Lastly the probability distribution of total coincidence counts $%
p_{c,x_{a}y_{b}}(n)$ is obtained by
\begin{eqnarray*}
p_{c,x_{a}y_{b}}(k) &=&\sum_{m,n=0}^{\infty }\delta
_{k,m+n}\,p_{p,x_{a}y_{b}}(m)\,p_{Acc,x_{a}y_{b}}(n)= \\
&=&\left( \,\lambda _{c,x_{a}y_{b}}t\right) ^{n}\,\frac{\exp
(-\lambda _{c,x_{a}y_{b}}t)}{n!},
\end{eqnarray*}
where the mean rate of total coincidence measured by an arbitrary pair of
$x_{a}$ and $%
y_{b}$ detectors is
\begin{equation}
\lambda _{c,x_{a}y_{b}}=\lambda _{p,x_{a}y_{b}}+\lambda
_{Acc,x_{a}y_{b}}. \label{lambdac}
\end{equation}

To characterize a particular QCKD procedure we embody the effect of the
transformations $%
\widehat{T}_{a}$ and $\widehat{T}_{b}$ on photon polarization by the
rotation matrices
\begin{equation}
\widehat{T}_{z}=\left(
\begin{array}{cc}
\cos \theta _{z} & \sin \theta _{z} \\
\sin \theta _{z} & -\cos \theta _{z}
\end{array}
\right) .  \label{ti2}
\end{equation}
We rewrite Eq. (\ref{lambdac}) in terms of the rotation angles $\theta
_{a},\theta
_{b},$ induced by transformations $\widehat{T}_{a}$ and $%
\widehat{T}_{b}$ on the polarization state of photons, by
replacing $p(x_{a},y_{b})$ with $p_{\theta _{a},\theta
_{b}}(x_{a},y_{b}),$ whose complete expression is in Appendix A$.$
More specifically, the calculated mean coincident counts are
made explicit in terms of angular settings $\theta _{a}$ and
$\theta _{b}$ as $\lambda _{c,x_{a}y_{b}}(\theta _{a},\theta
_{b})t.$

\section{\label{sec:level1} evaluation of QBER}

To characterize a particular QCKD procedure and to assess its
advantages, we evaluate particular quantities such as the QBER and
the sifted key for different types of QCKD protocols so far
experimentally implemented, i.e. BB84 protocol and Ekert's
protocols based on CHSH and Wigner's inequalities, respectively.

The QBER is a parameter for describing the signal quality in the
transmission of the sifted key, defined as the relative frequency
of errors induced by accidental coincidences, i.e. the number of
errors divided by the total size of the cryptographic sifted key
($K$) \cite{josab2}.

In other words, the QBER is given by total coincidence
provided by those detectors ``wrongly'' firing in coincidence
according to the chosen protocol. In fact the protocol establishes
which pair of detectors should fire to contribute to the key.

\subsection{\label{sec:level2} BB84 protocol}

Here we examine the BB84 protocol variant proposed for entangled
states in ref. \cite{bb92}. Recall that Alice and Bob
measure photons randomly and independently between two bases of
orthogonal quantum states. One basis corresponds to horizontal and
vertical linear polarization ($\oplus $), while the other to
linear polarizations rotated by 45${{}^{\circ }}$ ($\otimes $).
Only half of the photon pairs can contribute to the sifted key, as
only the subset of measurements performed with the two analyzers
in the same basis contributes.

The sifted key is given by
\begin{eqnarray}
K_{\text{BB84}}(\theta _{a}) &=&f_{basis}f_{setting}  \label{k1} \\
&&\sum_{x_{a}y_{b}}\left[
\begin{array}{c}
\lambda _{c,x_{a}y_{b}}(\theta _{a},\theta _{a})+ \\
\lambda _{c,x_{a}y_{b}}(\theta _{a}+\pi /4,\theta _{a}+\pi /4)
\end{array}
\right] t,  \notag
\end{eqnarray}
where $f_{basis}=1/2$ is the probability to measure in the right basis ($%
(\theta _{a},\theta _{a})$ and $(\theta _{a}+\pi /4,\theta
_{a}+\pi /4)$), while $f_{setting}=1/2$ is the probability to
measure in a particular analyzer setting ($(\theta _{a},\theta
_{a})$ or $(\theta _{a}+\pi /4,\theta _{a}+\pi /4)$).

All detectors contribute to the sifted key $K$, but only coincidences
between $1_{a}2_{b}$ and $%
2_{a}1_{b}$ correspond to the expected anticorrelation when
measurements are performed in the same basis ($\theta _{a}=\theta
_{b}$), while QBER contributions come from the coincidences between
detectors $1_{a}1_{b}$ and $%
2_{a}2_{b}$ (as it is clear from Eq.s (\ref{a1}) in Appendix A).
Therefore the QBER$_{\text{BB84}}$ explicit formula is
\begin{equation}
\text{QBER}_{\text{BB84}}(\theta
_{a})\text{=}\frac{\sum_{x=1,2}\left[
\begin{array}{c}
\lambda _{c,x_{a}x_{b}}(\theta _{a},\theta _{a})+ \\
\lambda _{c,x_{a}x_{b}}(\theta _{a}+\pi /4,\theta _{a}+\pi /4)
\end{array}
\right] t}{4 K_{\text{BB84}}(\theta _{a})}.  \label{qber1}
\end{equation}

To test the behavior of QBER, we simulate a realistic experiment
with parameters $\eta _{x_{a}}=\eta _{y_{b}}=0.5$ (quantum
efficiency of the four detectors)$,$ $\tau _{x_{a}}=\tau
_{y_{b}}=0.1$ (transmittance of the four channels)$,$ $\lambda
_{d,x_{a}}=\lambda _{d,y_{b}}=50$ s$^{-1}$ (dark count rate of the
four detectors), $D_{a}=D_{b}=100$ ns (total dead time of the
Alice's and Bob's detection systems), $\theta _{a}=0$ and $w=4$
ns. The entanglement parameters are $\epsilon =0.95$ and $\zeta
=1,$ the correlation level in the Alice channel is $\alpha
_{a}=0.25$ ($\protect\alpha _{a}=\protect\lambda
_{p}/(\protect\lambda _{p}+\sum_{x_{a}}\protect\lambda
_{u,x_{a}})$), and the correlated photon rate is
$\lambda_{p}=700\,$KHz.

\begin{figure}[tbp]
%[htbp]
\par
\begin{center}
\includegraphics[angle=0, width=8 cm, height=5 cm]{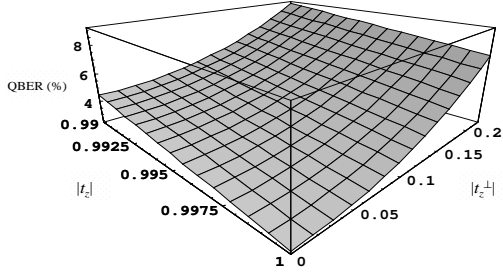}
\end{center}
\caption{ QBER for the BB84 protocol versus the PBS's coefficients
$\left| t_{z}\right| $ and $\left| t_{z}^{\perp }\right| .$}
\label{Figure 3}
\end{figure}

In Fig. 3 we show the dependence of QBER versus the optical
properties of real PBS, i.e. the transmittance $\left|
t_{z}\right| $ and $\left| t_{z}^{\perp }\right| $ ($z=a,b$) for
the states $\left| H_{z}\right\rangle $ and $\left|
V_{z}\right\rangle $ respectively. Results show how strongly the QBER
can be affected by the optical properties of PBSs, whose influence
has been neglected so far.

\begin{figure}[tbp]
%[htbp]
\par
\begin{center}
\includegraphics[angle=0, width=8 cm, height=5 cm]{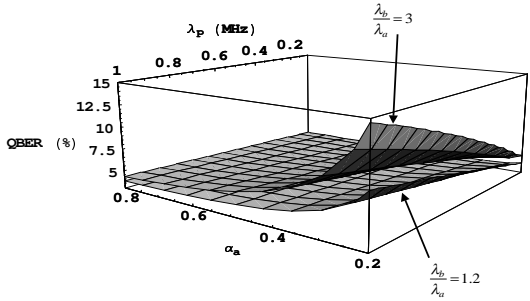}
\end{center}
\caption{ QBER for the BB84 protocol versus the correlation level
in the Alice channel $\protect\alpha _{a}=\protect\lambda
_{p}/(\protect\lambda _{p}+\sum_{x_{a}}\protect\lambda
_{u,x_{a}})$ and the correlated photon rate $\protect\lambda _{p}$
for two noise levels in the
channels ($\protect\lambda _{b}/\protect\lambda _{a}=1.2$ and $3$, where $%
\protect\lambda _{a}=\protect\lambda
_{p}+\sum_{x_{a}}\protect\lambda
_{u,x_{a}}$ and $\protect\lambda _{b}=\protect\lambda _{p}+\sum_{y_{b}}%
\protect\lambda _{u,y_{b}}$).} \label{Figure 4}
\end{figure}

In Fig. 4 the behavior of the QBER is presented versus the level
of correlation, $\alpha _{a}$, in the Alice channel and the rate,
$\lambda _{p}$, of the correlated photon pairs for two different
noise levels, i.e., the ratio between the mean rate of photons in
the two channels. Far from ideal conditions, the presence of
uncorrelated events in the two channels induces a non-linear
increase of the QBER. The other parameters are set as in Fig. 3
for realistic PBSs parameters, $\left| t_{z}\right| ^{2}=0.99 $ and
$\left| t_{z}^{\perp }\right| ^{2}=0.025$.

\subsection{\label{sec:level2} Ekert's protocol}

Ekert's protocol has the peculiarity of relying
on the completeness of quantum mechanics for security. Therefore, the possible
combined choices between Alice and Bob for analyzer settings split
into three groups: the first for key distribution, the second
containing the security proof, and the third garnering the
discarded measurements.

Here we consider two possible variants of Ekert's protocol: the
variant based on the Clauser-Horne-Shimony-Holt inequality (CHSH),
similar to the one proposed in Ref. \cite{qk3}, and the variant
based on Wigner's inequality \cite{qk2}.

\subsubsection{\label{sec:level3} Ekert's protocol based on CHSH inequalities%
}

To increase the number of measurements devoted to the key
distribution we consider the case where Alice and Bob measure
randomly among four analyzer settings and use the CHSH inequality to
test eavesdropping. In
this scheme, Alice's choices for the analyzer settings are  $%
\theta _{a}=(\theta _{a},\theta _{a}+\pi /8,\theta _{a}+\pi
/4,\theta _{a}+3\pi /8)$ and Bob's are $\theta _{b}=(\theta
_{a}+\pi /8,\theta _{a}+\pi /4,\theta _{a}+3\,\pi /8,\theta
_{a}+\pi /2)$.

The key distribution is performed when Alice and Bob's settings are
the same or orthogonal, so that the sifted key is
\begin{eqnarray*}
K_{\text{CHSH}}(\theta _{a}) &=&f_{setting} \\
&&\sum_{x_{a}y_{b}}\left[
\begin{array}{c}
\lambda _{c,x_{a}y_{b}}(\theta _{a},\theta _{a}+\pi /2)+ \\
\lambda _{c,x_{a}y_{b}}(\theta _{a}+\pi /8,\theta _{a}+\pi /8)+ \\
\lambda _{c,x_{a}y_{b}}(\theta _{a}+\pi /4,\theta _{a}+\pi /4)+ \\
\lambda _{c,x_{a}y_{b}}(\theta _{a}+3\pi /8,\theta _{a}+3\pi /8)
\end{array}
\right] t,
\end{eqnarray*}
where ($(\theta _{a},\theta _{a}+\pi /2),\,(\theta _{a}+\pi
/8,\theta _{a}+\pi /8),\,(\theta _{a}+\pi /4,\theta _{a}+\pi
/4),(\theta _{a}+3\pi
/8,\theta _{a}+3\pi /8))$ are angular settings generating the key and  $%
f_{setting}=1/16.$

In a maximally entangled state configuration, the detectors
contributing to the key should be $1_{a}1_{b}$ and $2_{a}2_{b}$
for the orthogonal analyzer settings and $1_{a}2_{b}$ and
$2_{a}1_{b}$ for the parallel settings. \ Thus the QBER is
calculated according to
\begin{widetext}
\begin{equation*}
\text{QBER}_{\text{CHSH}}(\theta _{a})\text{=}\frac{1}{4K_{\text{CHSH}%
}(\theta _{a})}\left\{ \sum_{x_{a},y_{b}(x\neq y)}\lambda
_{c,x_{a}y_{b}}(\theta _{a},\theta _{a}+\pi /2)t+\sum_{x_{a},y_{b}(x=y)}%
\left[
\begin{array}{c}
\lambda _{c,x_{a}y_{b}}(\theta _{a}+\pi /8,\theta _{a}+\pi /8)+ \\
\lambda _{c,x_{a}y_{b}}(\theta _{a}+\pi /4,\theta _{a}+\pi /4)+ \\
\lambda _{c,x_{a}y_{b}}(\theta _{a}+3\pi /8,\theta _{a}+3\pi /8)
\end{array}
\right] t\right\} ,
\end{equation*}
\end{widetext}
where the intuitive notation $%
\sum_{x_{a},y_{b} (x\neq y)}$ indicates the sum over $1_{a}2_{b}$
and $2_{a}1_{b}$
detectors and $\sum_{x_{a},y_{b} (x=y)}$ indicates the sum over
$1_{a}1_{b}$ and $%
2_{a}2_{b}.$

\subsubsection{\label{sec:level1} Ekert's protocol based on Wigner's
inequality}

As in Ref. \cite{qk2}, we consider the case of the Ekert's variant
where the security of the quantum channels follows from Wigner's
inequality. In this case, Alice and Bob measure randomly among four
analyzers settings whose choices are $\theta
_{a}=(\theta _{a}-\pi /6,\theta _{a})$ for Alice and
$\theta _{b}=(\theta _{a},\theta _{a}+\pi /6)$
 for Bob.
The key distribution is performed when Alice's and Bob's settings
are the same so that the sifted key is
\begin{equation*}
K_{\text{WI}}(\theta _{a})=f_{setting}\sum_{x_{a}y_{b}}\lambda
_{c,x_{a}y_{b}}(\theta _{a},\theta _{a})t,
\end{equation*}
with $f_{setting}=1/4.$

The QBER is calculated according to
\begin{equation*}
\text{QBER}_{\text{WI}}(\theta
_{a})\text{=}f_{setting}\frac{\left[ \lambda
_{c,1_{a}1_{b}}(\theta _{a},\theta _{a})+\lambda
_{c,2_{a}2_{b}}(\theta _{a},\theta _{a})\right]
t}{K_{\text{WI}}(\theta _{a})},
\end{equation*}
by  taking the detectors contributing to the wrong bits as
$1_{a}1_{b}$ and $2_{a}2_{b}.$

\begin{figure}[tbp]
%[htbp]
\par
\begin{center}
\includegraphics[angle=0, width=8 cm, height=5 cm]{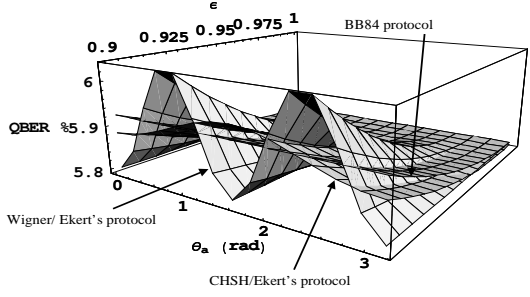}
\end{center}
\caption{ QBER in the case of CHSH and Wigner's inequality based
Ekert's
protocols together with BB84 protocol versus the angular analyzers settings $%
\protect\theta _{a}$ and the entanglement parameter
$\protect\epsilon $. The
parameters settings are $\protect\lambda _{a}=2.8$ $10^{6}$ s$^{-1},$%
\thinspace $\protect\alpha _{a}=0.25,$ $\protect\lambda
_{b}/\protect\lambda
_{a}=1.2,$ $\protect\eta _{x_{a}}=\protect\eta _{y_{b}}=0.5,$ $\protect\tau %
_{x_{a}}=\protect\tau _{y_{b}}=0.1,$ $\protect\lambda _{d,x_{a}}=\protect%
\lambda _{d,y_{b}}=50$ s$^{-1}$, $D_{a}=D_{b}=100$ ns, $w=4$ ns and $\protect%
\zeta =1$, taken from typical and experimental realistic values so
far implemented. PBSs are considered real with $\left|
t_{z}\right| ^{2}=0.99$ and $\left| t_{z}^{\perp }\right| ^{2}=0.025$ ($%
z=a,b $).} \label{Figure 5}
\end{figure}

In Fig. 5 we present a comparison of QBER levels for the BB84
protocol and the Ekert protocols considering both CHSH and
Wigner's inequality versus the analyzers angular setting $\theta
_{a}$ and the entanglement parameter $\epsilon $. Experimental
conditions are the same (low noise) as for Figs. 3 and 4. Results
highlight that
the QBER is sensitive to the angle $%
\theta _{a}$ when the ideal entanglement is not achieved for both
variants of Ekert's protocols. In the Wigner's case, the
sensitivity is so remarkable that this protocol has to be
considered less robust than BB84 and the CHSH based Ekert's
protocol.

\section{\label{sec:level1} Security and error correction}

The security of the BB84 variant protocol is based on a public
comparison between Alice and \ Bob's measurements on a
sufficiently large random subset of the sifted key, e.g. more than
half is recommended in \cite{bb92}.

The security proof for the CHSH-inequality-based Ekert's protocol
is evaluated with the specific choices of settings by the CHSH
inequality,
\begin{eqnarray*}
S(\theta _{a}) &=&E(\theta _{a},\theta _{a}+\pi /8)-E(\theta
_{a},\theta
_{a}+3\pi /8)+ \\
&&E(\theta _{a}+\pi /4,\theta _{a}+\pi /8)+E(\theta _{a}+\pi
/4,\theta
_{a}+3\pi /8), \\
S^{\prime }(\theta _{a}) &=&E(\theta _{a}+\pi /8,\theta _{a}+\pi
/4)-E(\theta _{a}+\pi /8,\theta _{a}+\pi /2)+ \\
&&E(\theta _{a}+3\pi /8,\theta _{a}+\pi /4)+E(\theta _{a}+3\pi
/8,\theta _{a}+\pi /2),
\end{eqnarray*}
where we have
\begin{eqnarray*}
E(\theta _{a},\theta _{b}) &=&M_{1_{a}1_{b}}(\theta _{a},\theta
_{b})-M_{1_{a}2_{b}}(\theta _{a},\theta _{b})+ \\
&&M_{2_{a}2_{b}}(\theta _{a},\theta _{b})-M_{2_{a}1_{b}}(\theta
_{a},\theta _{b}).
\end{eqnarray*}
Here
\begin{equation*}
M_{x_{a}y_{b}}(\theta _{a},\theta _{b})=\frac{\lambda
_{c,x_{a}y_{b}}(\theta _{a},\theta _{b})}{\sum_{x_{a}y_{b}}\lambda
_{c,x_{a}y_{b}}(\theta _{a},\theta _{b})}
\end{equation*}
is the normalized coincidence rate as a function of the analyzer
settings and detector choices. The terms $M_{x_{a}y_{b}}(\theta
_{a},\theta _{b})$ are commonly stored when experiments are
performed.

For maximally entangled states we have $\left| S_{\text{ql}%
}\right| =\left| S_{\text{ql}}^{\prime }\right| =2\sqrt{2}$, while
for any realistic local theory we have $\left| S,S^{\prime
}\right| \leq 2. $ It is expected that the presence of an
eavesdropper will reduce
the observed value of $\left| S,S^{\prime }\right| ,$ giving $\left| S_{%
\text{eve}},S_{\text{eve}}^{\prime }\right| \leq \sqrt{2},$ when
the eavesdropper measures photons over either one or both (total
eavesdropping) of Alice's and Bob's channels \cite{ekert}.

In the case of Wigner's-inequality-based Ekert's protocol, the
Wigner's inequality result has
\begin{eqnarray*}
W(\theta _{a}) &=&M_{1_{a}1_{b}}(\theta _{a}-\pi /6,\theta
_{a})+M_{1_{a}1_{b}}(\theta _{a},\theta _{a}+\pi /6)+ \\
&&-M_{1_{a}1_{b}}(\theta _{a}-\pi /6,\theta _{a}+\pi /6),
\end{eqnarray*}
giving for the maximally entangled states, $W_{\text{ql}}=-1/8$ and
$W\geq 0$ for any local realistic theory. As for the CHSH inequality,
it can be proved that the limit becomes $W_{\text{eve}}\geq 1/16$
for Eve detecting only one photon of the pair, while in the case
of total eavesdropping there is no boundary condition \cite{pc} .

\begin{figure}[tbp]
%[htbp]
\par
\begin{center}
\includegraphics[angle=0, width=8 cm, height=5 cm]{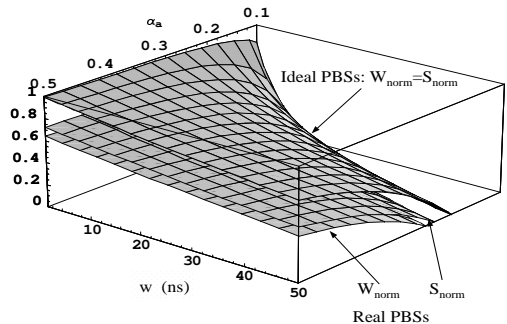}
\end{center}
\caption{ The CHSH and Wigner's inequalities parameters
S$_{\text{norm}}$ and W$_{\text{norm}}$ versus the
coincidence window $w$ and the correlation level in the Alice channel $%
\protect\alpha _{a}$. In the case of maximally entangled states we
have W$_{\text{norm}}=$S$_{\text{norm}}=$1, while the lower cut is
defined by the eavesdropping limit W$_{\text{norm}}\leq 0$ and
S$_{\text{norm}}\leq 0$ . The choice of parameter values are the
same of Fig. 4 except for $\protect\theta _{a}=0$ and
$\protect\epsilon =0.95$. The lower two surfaces represented
correspond to W$_{\text{norm}}$ and S$_{\text{norm}}$ in the case
of real PBSs, with parameters $\left| t_{z}\right| ^{2}=0.98$ and
$\left|
t_{z}^{\perp }\right| ^{2}=0.05.$ The higher surface represents S$_{\text{%
norm}}$=W$_{\text{norm}}$ in the case of ideal PBSs.}
\label{Figure 6}
\end{figure}

In Fig. 6 we compare the behaviors of the CHSH and Wigner's inequality
parameters, $S_{%
\text{norm}}$ =$\left( \left| S\right| -\left|
S_{\text{eve}}\right| \right) /\left( \left| S_{\text{ql}}\right|
-\left| S_{\text{eve}}\right| \right) $
and $W_{\text{norm}}=\left( W-W_{\text{eve}}\right) /\left( W_{\text{ql}}-W_{%
\text{eve}}\right) ,$ versus the coincidence window $w$ and the
correlation level in Alice's channel, $\alpha _{a}.$ The lower
surfaces represent the case of real PBSs, where $W_{\text{norm}}<$ $S_{\text{%
norm}}$, while the upper surface corresponds to \ ideal PBSs, where $W_{%
\text{norm}}=$ $S_{\text{norm}} $. We observe that, given the same
noise level in the system and real PBSs, Wigner's parameter
reaches the eavesdropping limit faster than the CHSH one does, revealing
the intrinsic weakness of Wigner's test against experimental
parameters.

Furthermore, the Wigner's security test guarantees against
eavesdropping strategies only for the detection of one photon of
the pair, while the CHSH security is independent on the adopted
strategy (see refs. \cite{ekert,pc}).

A satisfactory protocol must be able to recover from noise as well
as from partial leakage, allowing Alice and Bob to reconcile the
two strings of bits measured and distill from the sifted key a
corrected key. A strong need for the application of any
error-correction method is an \textit{a priori} knowledge of the
QBER, which provides information regarding how many times the
error-correction procedure must be applied to reduce the QBER to a
certain agreed level, commonly 1 \%. Here, we show an example of
error correction on an \textit{a priori} evaluated QBER according
to a common approach reported in Ref. \cite{qk2}, to
show that our model allows for prediction of the corrected key
length.

In general, Alice and Bob cannot distinguish between errors caused
either by an eavesdropper or by the environment. Thus, they must
assume that all errors are due to an eavesdropper and evaluate the
leaked information from the QBER. Also, even though by the
error-correction procedure one can disregard incorrect bits by
simply dropping them off in building the distilled key, the
residual knowledge of an eavesdropper may still not be faithfully
quantified by the reduced QBER obtained after the correction. The
effects of Eve's strategy is in fact equivalent to quantum noise
yielding eventually accidental coincidences, these last
contributing to both incorrect and correct bits transmitted, as it
is clear from Eqs. (\ref{k1}, \ref{qber1}). Hence the
error-correction procedure is not sufficient to cancel a potential
Eve's knowledge of part of the key due to accidental coincidence.

To prove this last assertion we introduce the quantum accidental
bit rate (QABR), a quantity related to accidental coincidences
and, in this sense, analogous to the QBER. However, the
error-correction procedure cannot reduce the QABR at the QBER
level. As an example, we give the QABR in the case of Ekert's
protocol based on Wigner's inequality,
\begin{equation*}
\text{QABR}_{\text{WI}}(\theta _{a})\text{=}f_{setting}\frac{%
\sum_{x_{a}y_{b}}\lambda _{Acc,x_{a}y_{b}}(\theta _{a},\theta _{a})t}{K_{%
\text{WI}}(\theta _{a})}.
\end{equation*}

\begin{figure}[tbp]
%[htbp]
\par
\begin{center}
\includegraphics[angle=0, width=8 cm, height=5 cm]{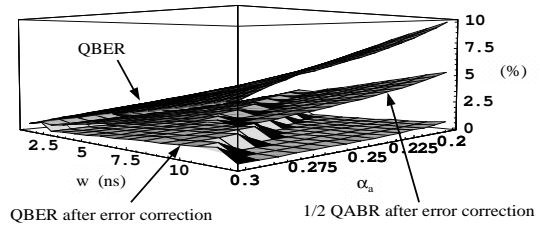}
\end{center}
\caption{  QBER in the case of BB84 protocol versus the
coincidence window $w$ and the correlation parameter in the Alice
channel $\alpha_{a}$: the sawtooth shape is due to the application
of the error correction procedure. The parameters are set as in
Figure 6 except that, in this case, ideal PBSs are considered. }
\label{Figure 7}
\end{figure}

Figure 7 shows the QBER with and without the application of the
error-correction procedure together with the 1/2 QABR vs the
coincidence window and the correlation parameter in Alice's
channel. The error-correction procedure is very ineffective at
reducing the QABR and consequently the possible effect of Eve's knowledge on
the corrected key.

\section{Conclusions}

This paper is concerned with an \textit{a priori} evaluation of
QCKD crucial parameters when entangled photons produced by SPDC
are exploited. The basic experimental feature consists in the
detection of coincident photons. Toward this aim, we developed a
statistical model to calculate the probability of accidental
coincidences contributing to errors in the sifted key, not
completely accounted for by simple experimental means.

We investigated the noise contribution due to imperfect source
generation and selection, imperfect polarizing beam splitters
performing polarization states analysis, and noisy and lossy
measurement system for photon-number detection. We emphasized some
basic system imperfections such as uncorrelated photons
collection, detection system deficiencies, detection system noise
due to detector dark counts, and electronic system
imperfections associated with non-ideal-time-correlation
measurements.

We discussed how this model can be adopted for the evaluation of
the QBER and the sifted key for different well-known protocols,
i.e., BB84 and Ekert's protocols based on both CHSH and Wigner's
inequality, and compared them to expected results.

Given that this model predicts precisely the QBER and the sifted
key, it ultimately guarantees a method to compare different
security criteria of the hitherto proposed QCKD protocols and
provides an objective assessment of performances and advantages of
different systems. Thus, it yields a method for an \textit{a
priori} evaluation of the tolerable experimental imperfections in
a practically implemented quantum system to establish the degree
of security and competitiveness of QCKD systems.

Finally, we used the model in a standard error-correction
procedure, observing that this does not completely cancel the
possible residual eavesdropping knowledge on the corrected key. We
emphasize that this model yields also the degree of security of
the corrected key, if a precise modelling of system imperfections
is provided.

Acknowledgment. This work was developed in collaboration with
Elsag S.p.A., Genova (Italy), within a project entitled ``Quantum
Cryptographic Key Distribution" co-funded by the Italian Ministry
of Education, University and Research (MIUR) - grant n. 67679/ L.
488. In addition Stefania Castelletto acknowledges the partial
support of the DARPA QuIST program.

\appendix

\section{\label{sec:level1} Interaction matrix $\widehat{U}_{%
\protect\psi \text{-PBS}}$}

We explicitly calculate the unitary transformation $\widehat{U}_{\psi \text{%
-PBS}}$ according to Eq.s (\ref{eq5}) obtaining a 16$\times $16
matrix
\begin{equation*}
\widehat{U}_{\psi \text{-PBS}}=\left(
\begin{array}{cccc}
U_{1} & U_{2} & U_{3} & U_{4} \\
U_{2} & U_{1} & U_{4} & U_{3} \\
U_{3} & U_{4} & U_{1} & U_{2} \\
U_{4} & U_{3} & U_{2} & U_{1}
\end{array}
\right) ,
\end{equation*}
where we indicate
\begin{widetext}
\begin{eqnarray*}
U_{1} &=&\left[
\begin{array}{cccc}
{\cal T}_{a}{\cal T}_{b} & 0 & 0 & 0 \\
0 & {\cal T}_{a}{\cal T}_{b}^{\perp } & 0 & 0 \\
0 & 0 & {\cal T}_{a}^{\perp }{\cal T}_{b} & 0 \\
0 & 0 & 0 & {\cal T}_{a}^{\perp }{\cal T}_{b}^{\perp }
\end{array}
\right] ;\;\;\;\;\;U_{2}=\left[
\begin{array}{cccc}
{\cal T}_{a}{\cal R}_{b} & 0 & 0 & 0 \\
0 & {\cal T}_{a}{\cal R}_{b}^{\perp } & 0 & 0 \\
0 & 0 & {\cal T}_{a}^{\perp }{\cal R}_{b} & 0 \\
0 & 0 & 0 & {\cal T}_{a}^{\perp }{\cal R}_{b}^{\perp }
\end{array}
\right] \\
U_{3} &=&\left[
\begin{array}{cccc}
{\cal R}_{a}{\cal T}_{b} & 0 & 0 & 0 \\
0 & {\cal R}_{a}{\cal T}_{b}^{\perp } & 0 & 0 \\
0 & 0 & {\cal R}_{a}^{\perp }{\cal T}_{b} & 0 \\
0 & 0 & 0 & {\cal R}_{a}^{\perp }{\cal T}_{b}^{\perp }
\end{array}
\right] ;\;\;\;\;\;\;\;\;U_{4}=\left[
\begin{array}{cccc}
{\cal R}_{a}{\cal R}_{b} & 0 & 0 & 0 \\
0 & {\cal R}_{a}{\cal R}_{b}^{\perp } & 0 & 0 \\
0 & 0 & {\cal R}_{a}^{\perp }{\cal R}_{b} & 0 \\
0 & 0 & 0 & {\cal R}_{a}^{\perp }{\cal R}_{b}^{\perp }
\end{array}
\right]
\end{eqnarray*}
\end{widetext}
and $\mathcal{T}_{z}=t_{z}/(t_{z}^{2}-r_{z}^{2})$, $\mathcal{R}%
_{z}=r_{z}/(t_{z}^{2}-r_{z}^{2})$ $(z=a,b),$ analogously for $\mathcal{T}%
_{z}^{\perp }$ and $\mathcal{R}_{z}^{\perp }.$

According to Eq.s (\ref{eqste}) and (\ref{ti2}) we obtain for $%
p_{\theta _{a},\theta _{b}}(x_{a},y_{b})$\ the following:
\begin{widetext}
\[
p_{\theta _{a},\theta _{b}}(1_{a},1_{b})=\frac{1}{1+\left|
\epsilon \right| ^{2}}\left\{
\begin{array}{c}
\cos ^{2}(\theta _{b})\left[
\begin{array}{c}
\cos ^{2}(\theta _{a})\left( \left|
\mathcal{T}_{a}\mathcal{T}_{b}^{\perp }\right| ^{2}+\left|
\mathcal{T}_{a}^{\perp }\mathcal{T}_{b}\epsilon \right|
^{2}\right) +
\sin ^{2}(\theta _{a})\left( \left| \mathcal{T}_{a}^{\perp }\mathcal{T}%
_{b}^{\perp }\right| ^{2}+\left|
\mathcal{T}_{a}\mathcal{T}_{b}\epsilon \right| ^{2}\right)
\end{array}
\right] + \\
\sin ^{2}(\theta _{b})\left[
\begin{array}{c}
\cos ^{2}(\theta _{a})\left( \left|
\mathcal{T}_{a}\mathcal{T}_{b}\right| ^{2}+\left|
\mathcal{T}_{a}^{\perp }\mathcal{T}_{b}^{\perp }\epsilon \right|
^{2}\right) +
\sin ^{2}(\theta _{a})\left( \left| \mathcal{T}_{a}^{\perp }\mathcal{T}%
_{b}\right| ^{2}+\left| \mathcal{T}_{a}\mathcal{T}_{b}^{\perp
}\epsilon \right| ^{2}\right)
\end{array}
\right] + \\
\sin (2\theta _{a})\cos (2\theta _{b})\mathrm{Re}(\epsilon \zeta
/2)\left(
\begin{array}{c}
-\left| \mathcal{T}_{a}\mathcal{T}_{b}\right| ^{2}+\left| \mathcal{T}%
_{a}^{\perp }\mathcal{T}_{b}\right| ^{2}+
\left| \mathcal{T}_{a}\mathcal{T}_{b}^{\perp }\right| ^{2}-\left| \mathcal{T}%
_{a}^{\perp }\mathcal{T}_{b}^{\perp }\right| ^{2}
\end{array}
\right)
\end{array}
\right\} ,
\]
\[
p_{\theta _{a},\theta _{b}}(1_{a},2_{b})=\frac{1}{1+\left|
\epsilon \right| ^{2}}\left\{
\begin{array}{c}
\cos ^{2}(\theta _{b})\left[
\begin{array}{c}
\cos ^{2}(\theta _{a})\left( \left|
\mathcal{T}_{a}\mathcal{R}_{b}^{\perp }\right| ^{2}+\left|
\mathcal{T}_{a}^{\perp }\mathcal{R}_{b}\epsilon \right|
^{2}\right) +
\sin ^{2}(\theta _{a})\left( \left| \mathcal{T}_{a}^{\perp }\mathcal{R}%
_{b}^{\perp }\right| ^{2}+\left|
\mathcal{T}_{a}\mathcal{R}_{b}\epsilon \right| ^{2}\right)
\end{array}
\right] + \\
\sin ^{2}(\theta _{b})\left[
\begin{array}{c}
\cos ^{2}(\theta _{a})\left( \left|
\mathcal{T}_{a}\mathcal{R}_{b}\right| ^{2}+\left|
\mathcal{T}_{a}^{\perp }\mathcal{R}_{b}^{\perp }\epsilon \right|
^{2}\right) +
\sin ^{2}(\theta _{a})\left( \left| \mathcal{T}_{a}^{\perp }\mathcal{R}%
_{b}\right| ^{2}+\left| \mathcal{T}_{a}\mathcal{R}_{b}^{\perp
}\epsilon \right| ^{2}\right)
\end{array}
\right] + \\
\sin (2\theta _{a})\cos (2\theta _{b})\mathrm{Re}(\epsilon \zeta
/2)\left(
\begin{array}{c}
-\left| \mathcal{T}_{a}\mathcal{R}_{b}\right| ^{2}+\left| \mathcal{T}%
_{a}^{\perp }\mathcal{R}_{b}\right| ^{2}+
\left| \mathcal{T}_{a}\mathcal{R}_{b}^{\perp }\right| ^{2}-\left| \mathcal{T}%
_{a}^{\perp }\mathcal{R}_{b}^{\perp }\right| ^{2}
\end{array}
\right)
\end{array}
\right\} ,
\]
\[
p_{\theta _{a},\theta _{b}}(2_{a},1_{b})=\frac{1}{1+\left|
\epsilon \right| ^{2}}\left\{
\begin{array}{c}
\cos ^{2}(\theta _{b})\left[
\begin{array}{c}
\cos ^{2}(\theta _{a})\left( \left|
\mathcal{R}_{a}\mathcal{T}_{b}^{\perp }\right| ^{2}+\left|
\mathcal{R}_{a}^{\perp }\mathcal{T}_{b}\epsilon \right|
^{2}\right) +
\sin ^{2}(\theta _{a})\left( \left| \mathcal{R}_{a}^{\perp }\mathcal{T}%
_{b}^{\perp }\right| ^{2}+\left|
\mathcal{R}_{a}\mathcal{T}_{b}\epsilon \right| ^{2}\right)
\end{array}
\right] + \\
\sin ^{2}(\theta _{b})\left[
\begin{array}{c}
\cos ^{2}(\theta _{a})\left( \left|
\mathcal{R}_{a}\mathcal{T}_{b}\right| ^{2}+\left|
\mathcal{R}_{a}^{\perp }\mathcal{T}_{b}^{\perp }\epsilon \right|
^{2}\right) +
\sin ^{2}(\theta _{a})\left( \left| \mathcal{R}_{a}^{\perp }\mathcal{T}%
_{b}\right| ^{2}+\left| \mathcal{R}_{a}\mathcal{T}_{b}^{\perp
}\epsilon \right| ^{2}\right)
\end{array}
\right] + \\
\sin (2\theta _{a})\cos (2\theta _{b})\mathrm{Re}(\epsilon \zeta
/2)\left(
\begin{array}{c}
-\left| \mathcal{R}_{a}\mathcal{T}_{b}\right| ^{2}+\left| \mathcal{R}%
_{a}^{\perp }\mathcal{T}_{b}\right| ^{2}+
\left| \mathcal{R}_{a}\mathcal{T}_{b}^{\perp }\right| ^{2}-\left| \mathcal{R}%
_{a}^{\perp }\mathcal{T}_{b}^{\perp }\right| ^{2}
\end{array}
\right)
\end{array}
\right\} ,
\]
\[
p_{\theta _{a},\theta _{b}}(2_{a},2_{b})=\frac{1}{1+\left|
\epsilon \right| ^{2}}\left\{
\begin{array}{c}
\cos ^{2}(\theta _{b})\left[
\begin{array}{c}
\cos ^{2}(\theta _{a})\left( \left|
\mathcal{R}_{a}\mathcal{R}_{b}^{\perp }\right| ^{2}+\left|
\mathcal{R}_{a}^{\perp }\mathcal{R}_{b}\epsilon \right|
^{2}\right) +
\sin ^{2}(\theta _{a})\left( \left| \mathcal{R}_{a}^{\perp }\mathcal{R}%
_{b}^{\perp }\right| ^{2}+\left|
\mathcal{R}_{a}\mathcal{R}_{b}\epsilon \right| ^{2}\right)
\end{array}
\right] + \\
\sin ^{2}(\theta _{b})\left[
\begin{array}{c}
\cos ^{2}(\theta _{a})\left( \left|
\mathcal{R}_{a}\mathcal{R}_{b}\right| ^{2}+\left|
\mathcal{R}_{a}^{\perp }\mathcal{R}_{b}^{\perp }\epsilon \right|
^{2}\right) +
\sin ^{2}(\theta _{a})\left( \left| \mathcal{R}_{a}^{\perp }\mathcal{R}%
_{b}\right| ^{2}+\left| \mathcal{R}_{a}\mathcal{R}_{b}^{\perp
}\epsilon \right| ^{2}\right)
\end{array}
\right] + \\
\sin (2\theta _{a})\cos (2\theta _{b})\mathrm{Re}(\epsilon \zeta
/2)\left(
\begin{array}{c}
-\left| \mathcal{R}_{a}\mathcal{R}_{b}\right| ^{2}+\left| \mathcal{R}%
_{a}^{\perp }\mathcal{R}_{b}\right| ^{2}+
\left| \mathcal{R}_{a}\mathcal{R}_{b}^{\perp }\right| ^{2}-\left| \mathcal{R}%
_{a}^{\perp }\mathcal{R}_{b}^{\perp }\right| ^{2}
\end{array}
\right)
\end{array}
\right\} .
\]
\end{widetext}

In the case of maximally entangled states, i.e. $\epsilon =1$ and $\zeta =1$%
, and ideal PBSs, i.e., $\left| \mathcal{R}_{z}\right| =\left| \mathcal{T}%
_{z}^{\perp }\right| =0$ and$\left| \mathcal{R}_{z}^{\perp
}\right| =\left| \mathcal{T}_{z}\right| =1,$ the
$\widehat{U}_{\psi \text{-PBS}}$ is simply given by
\begin{widetext}
\begin{eqnarray*}
U_{1} &=&\left[
\begin{array}{cccc}
1 & 0 & 0 & 0 \\
0 & 0 & 0 & 0 \\
0 & 0 & 0 & 0 \\
0 & 0 & 0 & 0
\end{array}
\right] ;\;\;\;\;\;\; U_{2}=\left[
\begin{array}{cccc}
0 & 0 & 0 & 0 \\
0 & i & 0 & 0 \\
0 & 0 & 0 & 0 \\
0 & 0 & 0 & 0
\end{array}
\right] ;\;\;\;\;\;\; U_{3}=\left[
\begin{array}{cccc}
0 & 0 & 0 & 0 \\
0 & 0 & 0 & 0 \\
0 & 0 & i & 0 \\
0 & 0 & 0 & 0
\end{array}
\right] ;\;\;\;\;\;\; U_{4}=\left[
\begin{array}{cccc}
0 & 0 & 0 & 0 \\
0 & 0 & 0 & 0 \\
0 & 0 & 0 & 0 \\
0 & 0 & 0 & 1
\end{array}
\right]
\end{eqnarray*}
\end{widetext}
and
\begin{eqnarray}
p_{\theta _{a},\theta _{b}}(1_{a},1_{b}) &=&p_{\theta _{a},\theta
_{b}}(2_{a},2_{b})=\sin ^{2}(\theta _{a}-\theta _{b})/2  \label{a1} \\
p_{\theta _{a},\theta _{b}}(1_{a},2_{b}) &=&p_{\theta _{a},\theta
_{b}}(2_{a},1_{b})=\cos ^{2}(\theta _{a}-\theta _{b})/2.  \notag
\end{eqnarray}

\section{\label{sec:level1} Dead-time correction determination}

According to Refs. \cite{madrid,lisbo}, in the case of
non-extending dead
time, the correction is $\pi _{z}=1/(1+\overline{n}_{z}D_{z}/t),$ where $%
\overline{n}_{z}$ are the mean number of photons counted in the
$z$ channel,
i.e. $\overline{n}_{z}=\sum_{x_{z}=1_{z},2_{z}}\sum_{k}k\,p_{\text{tot}%
,x_{z}}(k)$ with $p_{\text{tot},x_{z}}(k)$ calculated in the
absence of dead time $D_{z}$, and $\xi _{x_{z}}=\eta
_{x_{z}}\tau _{x_{z}}$. We can therefore write down the dead time
correction in this case as
\begin{equation*}
\pi _{a}=\left\{ 1+\sum_{x_{a}}\left[
\begin{array}{c}
\sum_{y_{b}}p(x_{a},y_{b})\,\eta _{x_{a}}\tau _{x_{a}}\lambda _{p}+ \\
\eta _{x_{a}}\tau _{x_{a}}\lambda _{u,x_{a}}+\lambda _{d,x_{a}}
\end{array}
\right] \,\,D_{a}\right\} ^{-1}
\end{equation*}
and
\begin{equation*}
\pi _{b}=\left\{ 1+\sum_{y_{b}}\left[
\begin{array}{c}
\sum_{x_{a}}p(x_{a},y_{b})\,\eta _{y_{b}}\tau _{y_{b}}\lambda _{p}+ \\
\eta _{y_{b}}\tau _{Y_{b}}\lambda _{u,y_{b}}+\lambda _{d,y_{b}}
\end{array}
\right] \,\,D_{b}\right\} ^{-1}
\end{equation*}
by noting that, when several devices are used in series, a good
approximation considers the whole apparatus to be a black box,
with a non-extending dead time equal to the largest of dead times
of the single
component \cite{madrid}. We showed in Ref. \cite{lisbo} that $\pi _{z\text{ }%
}$provides a satisfactory approximation for $t>>D_{z}$.


\begin{thebibliography}{99}
\bibitem{bennet&brassard}  C. Bennett and G. Brassard, in {\it{Proceedings of
the IEEE International Conference on Computers, Systems and Signal
Processing, Bangalore}}, (IEEE, New York 1984), p. 175.
\bibitem{ekert}  A. K. Ekert, Phys. Rev. Lett. {\bf 67}, 661 (1991).
\bibitem{bb92}  C. H. Bennett, G. Brassard, and N. D. Mermin, Phys. Rev.
Lett. {\bf 68}, 557
(1992).
\bibitem{brassard}  G. Brassard, N. Lutkenhaus, T. Mor, and B. C. Sanders,
Phys. Rev. Lett. {\bf
85}, 1330 (2000).
\bibitem{ekertrarity}  A. K. Ekert, J. G. Rarity, P. R. Tapster, and G. M.
Palma, Phys. Rev. Lett. {\bf 69}, 1293 (1992).
\bibitem{sasha}  A. V. Sergienko, M. Atature, Z. Walton, G. Jaeger, B. E. A.
Saleh, and M. C. Teich, Phys. Rev. A {\bf 60}, 2622 (1999).
\bibitem{qk2}  T. Jennewein, C. Simon, G. Weihs, H. Weinfurter, and A.
Zeilinger, Phys. Rev Lett. {\bf 84}, 4729 (2000).
\bibitem{qk3}  D. S. Naik, C. G. Peterson, A. G. White, A. J. Berglund, and
P. G. Kwiat, Phys. Rev. Lett. {\bf 84}, 4733
(2000).
\bibitem{qk1}  W. Tittel, J. Brendel, H. Zbinden, and N. Gisin, Phys.
Rev. Lett. {\bf 84}, 4737 (2000).
\bibitem{mandel}  L. Mandel, J. Opt. Soc. Am. B {\bf 1}, 108 (1984).
\bibitem{rarity0}  E. Jakeman and J. G. Rarity, Opt. Commun. {\bf 59}, 219
(1986).
\bibitem{rarity}  J. G. Rarity, P. R. Tapster, and E. Jakeman, Opt. Commun.
{\bf 62}, 201
(1987).
\bibitem{rarity2}  J. G. Rarity and P. R. Tapster, Appl. Phys. B {\bf 55}, 298
(1992).
\bibitem{prl}  D. C. Burnham and D. L. Weinberg,  Phys. Rev. Lett. {\bf
25}, 84 (1970).
\bibitem{klyshko}  D. N. Klyshko, {\it Photons and Nonlinear Optics},
(Gordon and Breach Science Publishers, New York, Amsterdam, 1988).
\bibitem{rarity4}  J. G. Rarity, K. D. Ridley, and P. R. Tapster, Appl.
Opt. {\bf 26,} 4616
(1987).
\bibitem{peninserg}  A. N. Penin and A. V. Sergienko, Appl. Opt. {\bf 30},
3582 (1991).
\bibitem{kwiat}  P. G. Kwiat, A. M. Steinberg, R. Y. Chiao, P. H.
Eberhard, and M. D. Petroff, Appl. Opt. {\bf 33}, 1844 (1994).
\bibitem{migdall}  A. L. Migdall , R. U. Datla, A. V. Sergienko, J. S.
Orszak, and Y. H. Shih , Metrologia {\bf 32}, 479 (1996).
\bibitem{metrolo}  S. Castelletto, A. Godone, C. Novero, and M. L.
Rastello, Metrologia {\bf 32}, 501 (1996).
\bibitem{teichSaleh}  M. M. Hayat, A. Joobeur, and B. E. A. Saleh, J. Opt.
Soc. Am. A {\bf 16}, 348 (1999).
\bibitem{josaB}  G. Brida , S. Castelletto, C. Novero, and M. L. Rastello,
J. Opt. Soc. Am. B {\bf 16}, 1623 (1999).
\bibitem{madrideta}  G. Brida, S. Castelletto, I. P. Degiovanni, C.
Novero, and M. L. Rastello, Metrologia {\bf 37}, 625 (2000).
\bibitem{rarity3}  P. R. Tapster, J. G. Rarity, and P. C. M. Owens, Phys.
Rev. Lett. {\bf 73},
1923 (1994).
\bibitem{gisin1}  W. Tittel, J. Brendel, H. Zbinden, and N. Gisin, Phys.
Rev. Lett. {\bf 81}, 3563 (1998).
\bibitem{zeil}  G. Weihs, T. Jennewein, C. Simon, H. Weinfurter, and A.
Zeilinger, Phys. Rev. Lett. {\bf 81}, 5039 (1998).
\bibitem{NC00} M. Nielsen and I. Chuang,
{\it Quantum Computation and Quantum Information}, (Cambridge
University Press, New York, 2000).
\bibitem{opteng}  N. Boeuf, D. Branning, I. Chaperot, E. Dauler, S. Guerin,
G. Jaeger, A. Muller, and A. Migdall, Opt. Eng. {\bf 39}, 1016
(2000).
\bibitem{kwiat2}  P. G. Kwiat, K. Mattle, H. Weinfurter, A. Zeilinger, A. V.
Sergienko, and Y. Shih, Phys. Rev. Lett. {\bf 75}, 4337 (1995).
\bibitem{foot1}  Entangled states may be
successfully generated for this purpose by two type I nonlinear
crystals, according to P. G. Kwiat, E. Waks, A. G. White, I.
Appelbaum, and P.H. Eberhard, Phys. Rev. A {\bf 60}, R773 (1999)%
, so that the general state is $\left| \psi \right\rangle =\left|
H_{a}\right\rangle \left| H_{b}\right\rangle -\exp [i\phi
]\,\left| V_{a}\right\rangle \left| V_{b}\right\rangle .$

\bibitem{scienceK}  P. G. Kwiat, A. J. Berglund, J. B. Altepeter, and A. G.
White, Science {\bf 290},
498 (2000).
\bibitem{berglund}  A. J. Berglund, arXiv:quantph/0010001 v2 (2000).
\bibitem{capri}  S. Castelletto, I. P. Degiovanni, and M. L. Rastello,
in {\it Quantum Communication, Computing, and Measurement 3,}
Proceedings of the Fifth International Conference on Quantum
Communication, Measurement and Computing, Capri, Italy, edited by
P. Tombesi and O. Hirota, (Kluwer Academic, New York, 2001), p.
131.
\bibitem{Perina}  J. Perina Jr., O. Haderka, and J. Soubusta, Phys. Rev. A
{\bf 64}, 052305
(2001).
\bibitem{josab2}  S. Castelletto, I. P. Degiovanni, and M. L. Rastello, J.
Opt. Soc. Am. B {\bf 19},
1247 (2002).
\bibitem{madrid}  S. Castelletto, I. P. Degiovanni, and M. L. Rastello,
Metrologia {\bf 37}, 613 (2000).
\bibitem{pc} S. Castelletto and
I. P. Degiovanni, \textit{private communication.}
\bibitem{lisbo}  S. Castelletto, I. P. Degiovanni, and M. L. Rastello, in
{\it Advanced Mathematical
and Computational tools in Metrology V}, edited by P. Ciarlini, M.
Cox, E. Filipe, F. Pavese, and D. Richter, (World Scientific
Company, Singapore, 2001), p. 41.




\end{thebibliography}
\end{document}